\documentclass[aps,prd,twocolumn,superscriptaddress,floatfix,nofootinbib,showpacs]{revtex4}

\usepackage{color}
\usepackage[dvips]{graphicx}
\usepackage{subfigure}

\newcommand{\lsim}{\mathrel{\mathop{\kern 0pt \rlap
  {\raise.2ex\hbox{$<$}}}
  \lower.9ex\hbox{\kern-.190em $\sim$}}}
\newcommand{\gsim}{\mathrel{\mathop{\kern 0pt \rlap
  {\raise.2ex\hbox{$>$}}}
  \lower.9ex\hbox{\kern-.190em $\sim$}}}

\newcommand{\be}{\begin{equation}}
\newcommand{\ee}{\end{equation}}
\newcommand{\beqarr}{\begin{eqnarray}}
\newcommand{\eeqarr}{\end{eqnarray}}

\begin{document}


\title{Light Neutralinos and WIMP direct searches}
\thanks{Preprint number: DFTT 16/2003}



%
\author{A. Bottino}
\affiliation{Dipartimento di Fisica Teorica, Universit\`a di Torino \\
Istituto Nazionale di Fisica Nucleare, Sezione di Torino \\
via P. Giuria 1, I--10125 Torino, Italy}

\author{F. Donato}
\affiliation{Dipartimento di Fisica Teorica, Universit\`a di Torino \\
Istituto Nazionale di Fisica Nucleare, Sezione di Torino \\
via P. Giuria 1, I--10125 Torino, Italy}

\author{N. Fornengo}
\affiliation{Dipartimento di Fisica Teorica, Universit\`a di Torino \\
Istituto Nazionale di Fisica Nucleare, Sezione di Torino \\
via P. Giuria 1, I--10125 Torino, Italy}

\author{S. Scopel}
\email{bottino@to.infn.it, donato@to.infn.it, fornengo@to.infn.it, scopel@to.infn.it}
\homepage{http://www.astroparticle.to.infn.it}
\affiliation{Dipartimento di Fisica Teorica, Universit\`a di Torino \\
Istituto Nazionale di Fisica Nucleare, Sezione di Torino \\
via P. Giuria 1, I--10125 Torino, Italy}

\date{\today}

\begin{abstract} \vspace{1cm}
The predictions of our previous analysis about 
possible low--mass ($m_{\chi} \lsim $ 50 GeV) 
 relic neutralinos are discussed in the light of some recent results from WIMP 
direct detection experiments.
It is proved that these light neutralinos are quite compatible with the new 
annual-modulation data of the DAMA Collaboration; 
our theoretical predictions are also compared with the upper 
bounds of the CDMS and EDELWEISS Collaborations.
\end{abstract}


\maketitle

Searches for neutralinos at colliders have not yet reached the
sensitivity required to place a direct lower bound on the neutralino
mass $m_{\chi}$. The commonly quoted and employed bound $m_{\chi}
\gsim 50$ GeV is derived from the lower bound on the chargino mass
determined at LEP2 ($m_{\chi}^{\pm} \gsim$ 100 GeV) under the
assumption that the $U(1)$ and $SU(2)$ gaugino masses $M_1$ and $M_2$
satisfy the standard relationship $M_1 \simeq \frac{1}{2} M_2$ at the
electroweak scale. This hypothesis is a consequence of the assumption
that these mass parameters have a common value at the grand
unification (GUT) scale.

In supersymmetric models with R-parity conservation and no
gaugino--unification assumption at the GUT scale, an absolute lower
limit on $m_{\chi}$ cannot be derived from the lower bound on the
chargino mass. Instead, it may be established by applying the upper
bound on the Cold Dark Matter (CDM) content in the Universe,
$\Omega_{CDM} \equiv \rho_{CDM}/\rho_c$, in combination with
constraints imposed on the Higgs and supersymmetric parameters by
measurements at colliders and other precision experiments (muon $g-2$,
$BR(b \rightarrow s + \gamma)$). This point was discussed in
Refs. \cite{1,2}, where a lower bound on the neutralino mass of about
6 GeV was established as a consequence of the recent 2$\sigma$
C.L. upper--limit $\Omega_{CDM} h^2 \leq 0.131$ obtained by the
analysis of the WMAP data \cite{spergel}.

In Refs. \cite{1,2} we also analyzed the properties of light
($m_{\chi} \lsim$ 50 GeV) relic neutralinos from the point of view of
their detection rates in WIMP direct search experiments. Two main
properties were pointed out: 1) direct detection rates for neutralinos
of mass lighter than 50 GeV reach levels within sensitivities of
current WIMP direct--detection experiments; 2) in the mass range 6 GeV
$\lsim m_{\chi} \lsim$ 25 GeV the detection rates are predicted to
fall in a very restricted range (this is at variance with what happens
at higher $m_{\chi}$, where the detection rates in 
an effective supersymmetric MSSM scheme (effMSSM)  vary over
decades \cite{35}). As shown in Refs. \cite{1,2} the property (2) is a
consequence the fact that the detection rate has a lower bound induced
by the upper limit on $\Omega_{CDM} h^2$.

Recalling that, for neutralino-matter interactions, coherent effects
systematically dominate over spin-dependent ones, the aforementioned
properties (1)--(2) are conveniently displayed in terms of the
quantity $\xi \sigma_{\rm scalar}^{(\rm nucleon)}$, where $\sigma_{\rm
scalar}^{(\rm nucleon)}$ is the neutralino--nucleon scalar
cross--section and $\xi$ is a rescaling factor between the neutralino
local matter density $\rho_{\chi}$ and the total local dark matter
density $\rho_0$: $\xi \equiv \rho_{\chi}/\rho_0$. Following a standard
assumption, $\xi$ may be taken as $\xi = {\rm min}(1,\Omega_\chi
h^2/(\Omega_{CDM} h^2)_{\rm min})$.

The supersymmetric model considered in the present paper is an effMSSM
scheme at the electroweak scale, with the following independent
parameters: $M_2, \mu, \tan\beta, m_A, m_{\tilde q}, m_{\tilde l}, A$
and $R \equiv M_1/M_2$. Notations are as follows: $\tan\beta$ the
ratio of the two Higgs v.e.v.'s: $\tan\beta\equiv$$<H_2^0>$/$<H_1^0>$,
$\mu$ is the Higgs mixing mass parameter, $m_A$ the mass of the CP-odd
neutral Higgs boson, $m_{\tilde q}$ is a soft--mass common to all
squarks, $m_{\tilde l}$ is a soft--mass common to all sleptons, $A$ is
a common dimensionless trilinear parameter for the third family,
$A_{\tilde b} = A_{\tilde t} \equiv A m_{\tilde q}$ and $A_{\tilde
\tau} \equiv A m_{\tilde l}$ (the trilinear parameters for the other
families being set equal to zero).  Since we are here interested in
light neutralinos, we consider values of $R$ lower than its standard
value: $R_{\rm GUT} \simeq 0.5$; for definiteness, we take $R$ in the
range: 0.01 - 0.5.  In the scanning of the supersymmetric parameter
space, we use the following ranges of the MSSM parameters: $1 \leq
\tan \beta \leq 50$, $100\, {\rm GeV }\leq |\mu|, M_2, m_{\tilde q},
m_{\tilde l} \leq 1000\, {\rm GeV }$, ${\rm sign}(\mu)=-1,1$, $90\,
{\rm GeV }\leq m_A \leq 1000\, {\rm GeV }$, $-3 \leq A \leq 3$.  We
impose the experimental constraints: accelerators data on
supersymmetric and Higgs boson searches, measurements of the $b
\rightarrow s + \gamma$ decay and of the muon anomalous magnetic
moment $a_\mu \equiv (g_{\mu} - 2)/2$ (the range $-160 \leq \Delta
a_{\mu} \cdot 10^{11} \leq 680 $ is used here for the deviation of the
current experimental world average from the theoretical evaluation
within the Standard Model).  We notice that to satisfy the $b
\rightarrow s + \gamma$ constraint is not trivial at small $m_A$
values, since it requires some cancellation between the contributions
due to the Higgs--quark and the chargino-squark loops.  However, it
turns out that the extent of compensation between the two terms is
limited to an effect of 30-60 $\%$. The range used here for the
branching ratio is $2.18 \times 10^{-4} \leq BR (b \rightarrow s +
\gamma) \leq 4.28 \times 10^{-4}$.  For $(\Omega_{CDM} h^2)_{\rm min}$
we use here the value $(\Omega_{CDM} h^2)_{\rm min} = 0.095$ \cite{2},
derived at the 2$\sigma$ C.L. from the analysis of
Ref. \cite{spergel}.

\begin{figure}
{\includegraphics[width=\columnwidth]{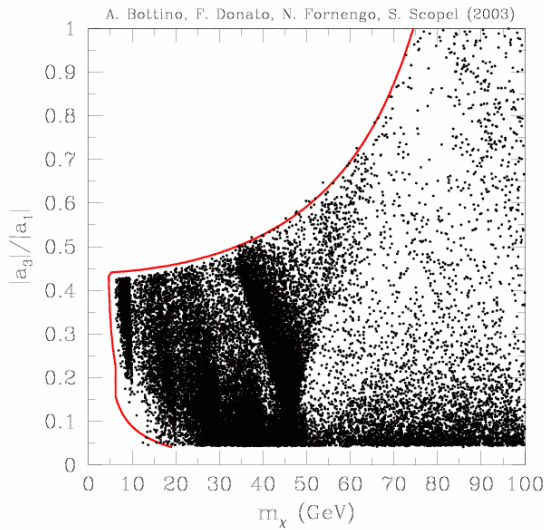}}
\caption{Ratio of the $\tilde{H}_1^\circ$ content to the $\tilde{B}$ content
in the neutralino composition
as a function of the neutralino mass $m_\chi$. The upper branch of 
the solid line derives from neutralino diagonalization (see text), while 
the lower one is obtained from 
$\Omega_{\chi} h^2 \leq (\Omega_{CDM} h^2)_{\rm max}=0.131$.
\label{fig:a3a1_mchi}}
\end{figure}

As discussed in Refs. [1-2], neutralino configurations at small $m_{\chi}$ have
a
dominant bino component with a small mixture with $\tilde{H}_1^0$, {\it i.e.}
writing the neutralino as $\chi \equiv a_1 \tilde B +
a_2 \tilde W^{(3)} + a_3 \tilde H_1^{\circ} + a_4 \tilde H_2^{\circ}$, 
one has $|a_1| >> |a_3| >> |a_2|, |a_4|$. In this regime
the ratio
$|a_3|/ |a_1|$ is given by the analytic expression
\begin{equation}
\frac{|a_3|}{|a_1|} \simeq \sin \theta_{W} \; \sin \beta \;
\frac{m_Z}{|\mu|} \lsim 0.42 \; \sin \beta,
\label{ratio}
\end{equation}
\noindent
where in the last step we have taken into account the experimental
lower bound $\mu \gsim $ 100 GeV.  The allowed range of the ratio
$|a_3|/|a_1|$ for increasing values of $m_{\chi}$ is displayed in
Fig. 1. The upper boundary line is given by intrinsic properties in
the diagonalization of the neutralino mass matrix and is obtained by
maximizing the parameters $M_2$ and $\tan \beta$ and by minimizing
$\mu$ (for very small $m_{\chi}$, Eq. (\ref{ratio}) applies). The
lower boundary line is derived from the condition $\Omega_{\chi} \leq
(\Omega_{CDM})_{max}$ (using Eq. (5) of Ref. [2]). These two boundary
lines fit well with the scatter plot obtained by a numerical scanning
of the supersymmetric parameter space, also displayed in Fig.
\ref{fig:a3a1_mchi}.  For $m_\chi \gsim 20$ GeV the cosmological bound
is satisfied by the $\tilde{\tau}$--exchange in the annihilation
cross--section, and $|a_3|$ is no longer constrained from below. The
lower bound on $|a_3|/|a_1|$ in the scatter plot of
Fig. \ref{fig:a3a1_mchi} is due to the upper value (1 TeV) in the
range of $\mu$ employed in our calculation.

\begin{figure*}
{\includegraphics[width=\columnwidth]{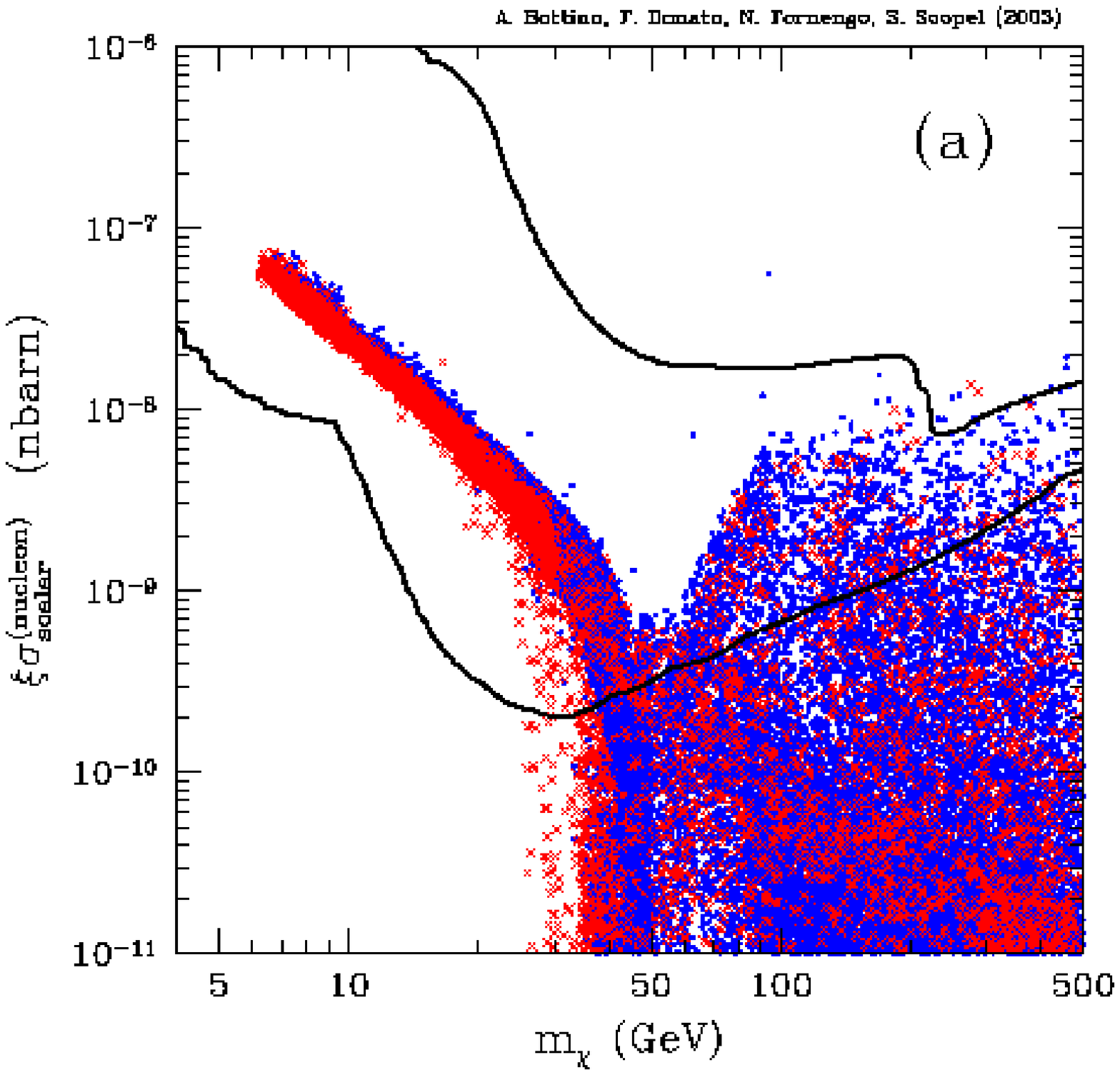}}
{\includegraphics[width=\columnwidth]{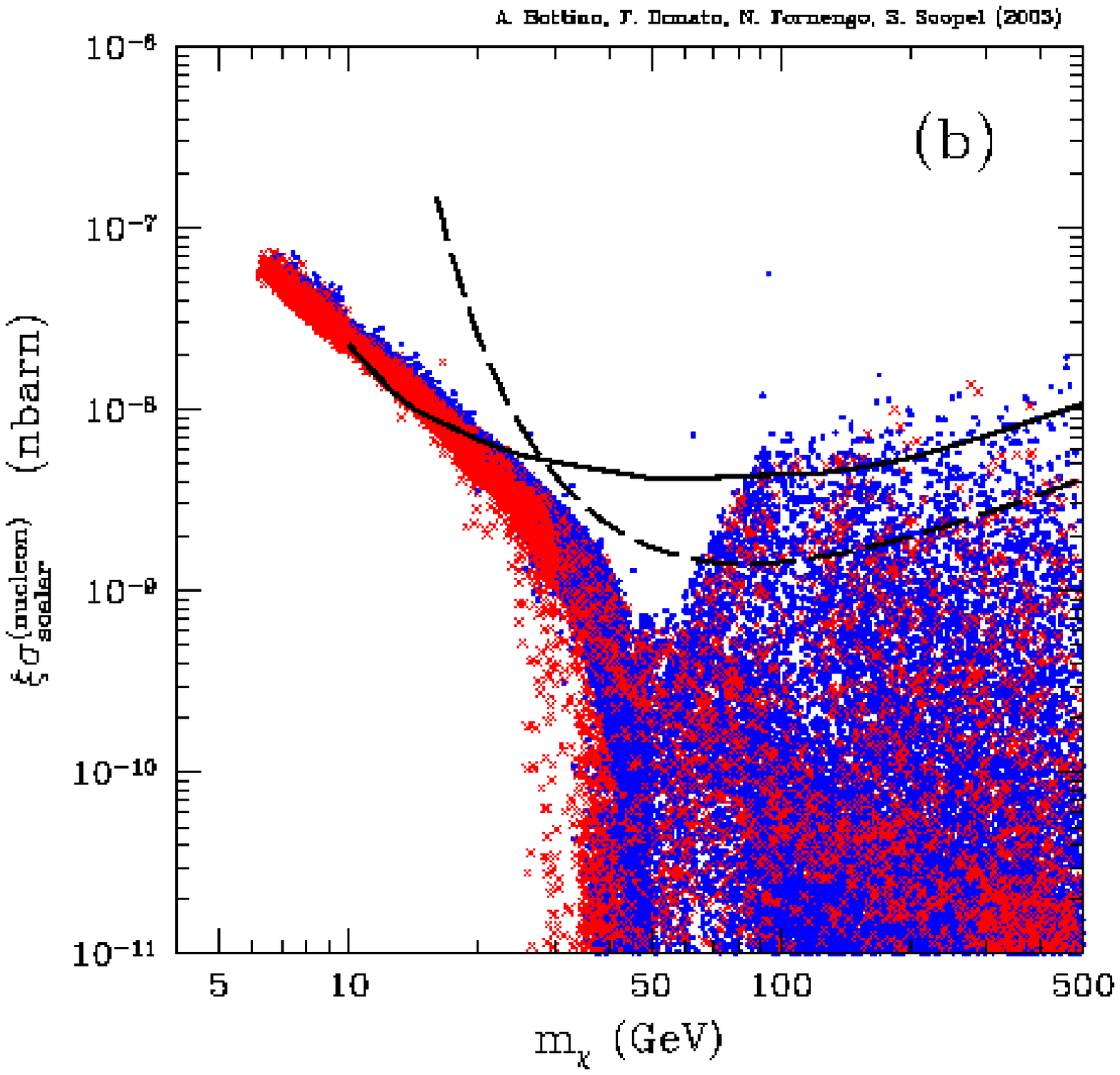}}
\caption{Scatter plot of $\xi \sigma_{\rm scalar}^{(\rm nucleon)}$
versus $m_{\chi}$.  Crosses (red) and dots (blue) denote neutralino
configurations with $\Omega_{\chi} h^2 \geq (\Omega_{CDM} h^2)_{\rm
min}$ and $\Omega_{\chi} h^2 < (\Omega_{CDM} h^2)_{\rm min}$,
respectively ($(\Omega_{CDM} h^2)_{\rm min}=0.095$) (a) The curves
delimit the DAMA region where the likelihood-function values are
distant more than $4 \sigma$ from the null (absence of modulation)
hypothesis \cite{dama2}; this region is the union of the regions
obtained by varying the WIMP DF over the set considered in Ref.
\cite{bcfs}. (b) The solid and the dashed lines are the experimental
upper bounds given by the CDMS \cite{cdms2} and the EDELWEISS
\cite{edelweiss} Collaborations, respectively, under the hypothesis
that the WIMP DF is given by an isothermal distribution with a
standard set of astrophysical parameters.
\label{fig:sigma_mchi}}
\end{figure*}

In Fig. \ref{fig:sigma_mchi} we show a scatter plot of the quantity
 $\xi \sigma_{\rm scalar}^{\rm (nucleon)}$ as a function of
 $m_{\chi}$. This scatter plot  shows that, in the mass
range 6 GeV $\lsim m_{\chi} \lsim$ 25 GeV, the quantity $\xi
\sigma_{\rm scalar}^{\rm (nucleon)}$ falls in a narrow funnel (see
property (2) above); this funnel is delimited from below by
configurations with $\Omega_{\chi} h^2 \sim (\Omega_{CDM} h^2)_{\rm
max}=0.131$, and delimited from above by supersymmetric configurations
with a very light Higgs boson (close to its lower experimental bound
of 90 GeV) and with an $\Omega_{\chi} h^2$ below $(\Omega_{CDM}
h^2)_{\rm min}$. For $m_{\chi} \lsim $ 10 GeV only values of 
$30 \lsim \tan \beta \leq 50$ and 100 GeV $ \leq |\mu| \lsim $ 300 GeV  
contribute, while in the interval 
10 GeV $\lsim m_{\chi} \lsim$ 25 GeV $\tan \beta$ extends also to lower values 
around 8 and $|\mu|$ is not significantly constrained.
Moreover, for  $m_{\chi} \lsim$ 20 GeV,  $m_{A}$ is strongly 
bounded from above 
by $(\Omega_{CDM} h^2)_{\rm max}$, as shown in Fig. 3 of Ref. \cite{2}. 
Notice that the dip at $\simeq $ 45 GeV is due to the Z--pole in the 
annihilation cross--section. 

It is also remarkable that, within the funnel, the
size of $\xi \sigma_{\rm scalar}^{\rm (nucleon)}$ is large enough to
make light relic neutralinos explorable by WIMP direct experiments
with the current sensitivities. To illustrate this point, let us turn
now to a comparison of our predictions with experimental data
\cite{dama2,cdms2} which became available after our analysis of
Refs. \cite{1,2} and with those of Ref. \cite{edelweiss}.

In Refs. \cite{1,2} we anticipated
that a detector with a high exposure and a low threshold such as DAMA
 \cite{dama1} might provide significant information, not only for
neutralinos with $m_{\chi} >$ 50 GeV, but also for neutralinos in the
 mass range 6 GeV $\lsim m_{\chi} \lsim$ 50 GeV. In Ref. \cite{2}
we could only give an estimate of the expected effects in the 
measurement of the annual--modulation variation performed 
by the DAMA Collaboration, since no analysis of the 
experimental data at low WIMP masses was available
at that time. 

Now, the recent presentation of new results by the DAMA Collaboration
\cite{dama2} allows us to compare directly our theoretical predictions
to actual experimental data. In fact, one has now the results of a
much larger exposure than in the past, about 108,000 kg $\cdot$ day
and, most important, an analysis of the full set of experimental data
in terms of a spin--independent effect over an unconstrained range for
the mass of a generic WIMP. The results of this analysis are reported
in Fig. \ref{fig:sigma_mchi}(a), where the contour line (after Fig.28 of Ref.
\cite{dama2}) delimits a region of the $m_{\chi}-\xi \sigma_{\rm
scalar}^{\rm (nucleon)}$ plane, where the likelihood-function values
are distant more than $4 \sigma$ from the null (absence of modulation)
hypothesis. In deriving this contour line, the DAMA Collaboration has
taken into account a rather large class of possible phase--space
distribution functions (DF) for WIMPs in the galactic halo. The
categories of DFs considered in Ref. \cite{dama2} are those analyzed
in Ref. \cite{bcfs}; the annual--modulation region displayed in
Fig. \ref{fig:sigma_mchi}(a) 
is the union of the regions obtained by varying over the set
of the DFs considered in Ref. \cite{bcfs}.  From Fig. \ref{fig:sigma_mchi}(a)
 we derive
that the entire population of relic neutralinos with $m_{\chi} \lsim$
25 GeV as well as a significant portion of those with a mass up to
about 50 GeV are within the annual--modulation region of the DAMA
Collaboration.  Thus, this yearly effect could be due to relic
neutralinos of light masses, in alternative to the other possibility
which we already discussed in Refs. \cite{35} on neutralinos with
masses above 50 GeV, and which is reconfirmed by the present analysis.

Another experiment of WIMP direct detection, run by the CDMS Collaboration, has recently 
published new data \cite{cdms2}. Their results are given in terms of an upper bound 
\footnote{We consider here only the upper bound derived by the CDMS
Collaboration without neutron subtraction. The upper limit obtained by
subtracting an estimated neutron background appears too
model--dependent; this point will be overcome, when the CDMS
experiment is run in an underground location, as foreseen by the
Collaboration.} on $\sigma_{\rm scalar}^{\rm (nucleon)}$ for a given
DF (an isothermal distribution) and for a single set of the
astrophysical parameters: $\rho_0 =$ 0.3 GeV $\cdot$ cm$^{-3}$, $v_0$
= 220 km $\cdot$ s$^{-1}$ ($v_0$ is the local rotational velocity).
This upper bound is displayed in Fig. \ref{fig:sigma_mchi}(b) together with our
theoretical scatter plot.  Thus, we see that {\it in case of an
isothermal DF with the representative values of parameters given
above}, a sizeable subset of supersymmetric configurations in the mass
range 10 GeV $\lsim m_{\chi} \lsim$ 20 GeV would be incompatible with
the experimental upper bound (together with some configurations with
$m_{\chi} \gsim$ 80 GeV).  However, this conclusion cannot be drawn in
general.  In fact, to set a solid constraint on the theoretical
predictions, it is necessary to derive from the experimental data the
upper bounds on $\xi \sigma_{\rm scalar}^{\rm (nucleon)}$ for a large
variety of DFs and of the corresponding astrophysical parameters (with
their own uncertainties); the intersection of these bounds would
provide an absolute limit to be used to possibly exclude a subset of
supersymmetric population. An investigation by the CDMS Collaboration
along these lines would be very interesting.

Among other experiments of WIMP direct detection, the EDELWEISS experiment has 
published an upper bound which somewhat approaches the region of the low--mass neutralino
population. This upper limit, again provided for a single DF  (the  
isothermal sphere with a standard set of astrophysical parameters)
 is also displayed in Fig. \ref{fig:sigma_mchi}(b); 
it turns out to be marginal for the low--mass population, 
since it is tangent to 
our supersymmetric scatter plot (at  $m_{\chi} \sim$ 30 GeV). The 
argument given before applies again in this case; one should vary the analytical forms of the DF, in order 
to derive 
a model--independent bound on $\xi \sigma_{\rm scalar}^{\rm (nucleon)}$. As for the low--mass 
configurations ($m_{\chi} \lsim$ 50 GeV), since the current upper limit is 
already marginal for the isothermal DF, one does not expect any model--independent constraint. 
 However, useful constraints could be derived for 
higher masses.

In conclusion, we have shown that the experimental exploration of the
low--mass neutralino population, theoretically analyzed in our papers
of Refs. \cite{1,2}, is already under way in case of some experiments
of WIMP direct detection and within the reach of further investigation in
the near future.

We have compared our predictions with available results of various
experiments separately, since the experimental results of different
Collaborations are not all derived under the same assumptions on the
WIMP phase--space distribution function.  A more effective comparison
of theoretical results with experimental data will be feasible, only
when the analysis of different experimental results in terms of
$m_{\chi}-\xi \sigma_{\rm scalar}^{\rm (nucleon)}$ is presented for
each analytic form of the DF, separately. This is also the unique way
of comparing results of different experiments among themselves.

We finally notice that, in direct detection experiments, lighter WIMPs
have to be faster, as compared to the heavier ones, in order to
deposit a recoil energy above the energy threshold.  As a consequence,
for light WIMPS the calculation of expected rates and the
determination of upper limits on the cross--section are very sensitive
to the value assigned to the escape velocity and, more generally, to
the details of the high--velocity tail of the DF.  This introduces an
important uncertainty, since for high--velocity WIMPS the assumption
of thermalization, which for instance is assumed in all the models
considered in the analysis of Ref. \cite{bcfs}, is less robust than
for the bulk of the distribution: non--thermal components, such as
streams, could have a sizeable or even dominant weight, affecting the
usual estimates for expected rates.

{\sc Note Added.}

After submission of the present paper, updated evaluations of the 
$e^+-e^-$--based and $\tau$--based lowest-order (LO)
hadronic polarization contribution to the muon magnetic moment 
have been presented by M. Davier et al. (hep-ph/0308213). 
In order to obtain a conservative range for the deviation 
of the standard model value of $a_\mu$ from its experimental 
determination, we consider the $e^+-e^-$--based LO hadronic 
contribution by K. Hagiwara et al. (hep-ph/0209187) together with
the $\tau$--based LO hadronic contribution by M. Davier et al.
Combining the evaluation by K. Hagiwara et al.  with the other standard 
model contributions, we find for the deviation from the world 
average experimental result the 2$\sigma$ range: 
$ 133 \leq  \Delta a_\mu \leq 569$. Instead, using
the $\tau$--based Davier et al. result, 
one obtains the 2$\sigma$ interval: $ -142 \leq  \Delta a_\mu \leq 286$.
If we combine conservatively these two determinations we finally have 
$ -142 \leq  \Delta a_\mu \leq 569$.
Employing this range instead of the one used 
in the derivation of the scatter plot in Fig. \ref{fig:sigma_mchi}
 does not modify 
 the features of the plot in any significant way.

\end{document}